# Localization of 5f-electrons and pressure effects on magnetism in U intermetallics in the light of spin-fluctuation theory

P. Opletal[1], J. Valenta[1,2], P. Proschek[1], V. Sechovský[1], J. Prokleška[1]

[1]Charles University, Faculty of Mathematics and Physics, Department of Condensed Matter Physics, Ke Karlovu 5, Prague 2, Czech Republic

[2]National Institute for Materials Science, Thermal Energy Materials Group, International Center for Materials Nanoarchitectonics (MANA), 1-2-1, Sengen, Tsukuba, Ibaraki 305-0047, Japan

**Abstract**

UCoGa and URhGa, two isostructural compounds, show opposite signs of the initial response of Curie temperature to applied hydrostatic pressure. To determine the physical origin of this difference the magnetization data measured with respect to temperature, magnetic field and hydrostatic pressure were analyzed in the framework of the Takahashi's spin-fluctuation theory. The parameters $T_0$ and $T_A$ characterizing the distribution widths of the spin-fluctuation spectrum in the energy and wave vector space, respectively, and $T_C/T_0$, the degree of the 5f-electron localization has been determined. Examination of available experimental data for the other U*TX* (*T* = a transition metal, *X* = Al, Ga) ferromagnets having the ZrNiAl-type structure revealed some correlations between the degree of the 5f-electron localization represented by the spin-fluctuation parameters and the response of Curie-temperature on the applied pressure. The extent to which this correlation can be used to describe the localization and magnetic behavior of other uranium ferromagnetic compounds is discussed.

**Introduction**

Nowadays, ferromagnetic quantum criticality is a heavily studied phenomenon. Hydrostatic pressure acting on the studied material has proven to be a suitable parameter for tuning magnetic state.

In the present study, we focus on a group of compounds from the U*TX* family (*T*- transition metal, *X*- p-metal) crystallizing in the hexagonal ZrNiAl structure. Many of these compounds are ferromagnetic with a relatively large span of Curie temperatures and magnitudes of ordered moments [1] and may therefore be used to study discontinuous phase transitions with variable parameters of magnetic order, yet with a fixed crystal structure. In several of these compounds the presence of a discontinuous phase transitions has been recently confirmed (e.g. UCoAl [2,3], URhAl [4], UCoGa [5]). On the other hand, two members of this group, UPtAl [6] and URhGa [7], respectively, show an initial (in the range of several GPa) increase of $T_C$ with increasing pressure. It is therefore desirable to be able to predict the pressure behavior of $T_C$ in these compounds.

In this paper, Takahashi's spin-fluctuation theory (TSFT) [8] is used to determine if an increasing $T_C$ with increasing pressure can be expected. In this theory, the total amplitude of the local spin fluctuations (SF) is constant as a function of temperature. This enables one to determine the value of $F_1$, the mode-mode coupling term as the coefficient of the $M^4$ term in the Landau expansion of the free energy

$$F_m(M) = F_m(0) + \frac{1}{2(g\mu_B)^2\chi}M^2 + \frac{F_1}{4(g\mu_B)^4 N_0^3}M^4,$$

and the values of $T_0$ and $T_A$, which represent the distribution widths of the SF spectrum in energy and wave vector space, respectively

$$\left(\frac{T_C}{T_0}\right)^{5/6} = \frac{M_S^2}{5g^2 C_{4/3}}\left(\frac{15cF_1}{2T_C}\right)^{1/2}, \quad (2)$$

$$\left(\frac{T_C}{T_A}\right)^{5/3} = \frac{M_S^2}{5g^2 C_{4/3}}\left(\frac{2T_C}{15cF_1}\right)^{1/3}, \quad (3)$$

where $g$ is the gyro-magnetic ratio, $\mu_B$ is the Bohr's magneton and $N_0$ is the Avogadro number, $M_s$ is the spontaneous magnetic moment, $C_{4/3}$ is constant (= 1.006089 ⋯) and $c = \frac{1}{2}$. The value of $F_1$ is determined from the slope of the Arrott plot ($M^2$ vs $M/H$) at low temperatures [8]. The ratio $T_C/T_0$ corresponds to the degree of localization of the electrons responsible for the magnetization and ranges from $T_C/T_0 = 1$ for the entirely localized case to $T_C/T_0 \to 0$ for completely delocalized [8].

The compounds targeted by the study are UCoGa, with $T_C = 48$ K [9,10] and $T_C$ decreasing with applied pressure [5], and URhGa with $T_C = 41$ K [10] and $T_C$ increasing with pressures up to 6 GPa [7]. The analysis of magnetization data observed at ambient pressure reveals a clear difference of the corresponding $F_1$, $T_0$, $T_A$ and $T_C/T_0$ values obtained for the two U*TX* compounds. A considerably higher degree of 5*f*-electron localization (higher $T_C/T_0$) in conjunction with the narrower SF spectrum both in energy and wave-vector space (smaller $T_0$ and $T_A$) have been documented for URhGa in comparison with UCoGa. The results of the analysis of pressure induced changes of $F_1$, $T_0$, $T_A$ and $T_C/T_0$ corroborate the proposed scenario of $dT_C/dP > 0$ for URhGa in contrast to $dT_C/dP < 0$ for UCoGa. The evolution of $dT_C/dP$ of the hexagonal U*TX* ferromagnets with the ZrNiAl-type structure including UCoGa and URhGa reveals correlation with $T_C/T_0$ ratio across this isostructural series. A discussion focused on a comparison with the corresponding data known for orthorhombic U*TX* ferromagnets and UGe$_2$ indicates that the above correlation does not have universal validity for all U ferromagnets.

**Experimental**

Single crystals of UCoGa and URhGa were prepared by the Czochralski method using a tri-arc furnace. For UCoGa, details of the single-crystal preparation and annealing are presented elsewhere [11]. URhGa was grown with a pulling speed of 12 mm/hr. The ingot was wrapped in tantalum foil and annealed at 900 °C in an evacuated quartz tube. Magnetization measurements at ambient and hydrostatic pressures were performed in a MPMS XL 7T magnetometer (Quantum Design). Since the U*TX* compounds crystallizing in the hexagonal ZrNiAl structure exhibit huge uniaxial magnetocrystalline anisotropy with the entire magnetic moment concentrated to the *c*-axis [1] only the magnetization data for this field direction were measured and used for the presented analysis. For pressure experiments, a small CuBe hydrostatic cell [12] was used, with Daphne 7373 oil as a pressure medium. The superconducting transition of Pb was used to determine the pressure in the cell at low temperatures.

**Results and Discussion**

The temperature dependencies of the magnetization of UCoGa and URhGa measured in a low applied field (0.1 T) at various pressures are shown in Figs. 1 and 2, respectively. The values of Curie temperature ($T_C$) were estimated as the temperature of the inflection point of these thermomagnetic curves. For UCoGa, $T_C$ decreases while for URhGa $T_C$ increases with increasing pressure in agreement with earlier high pressure studies [5,7]. The phase transition in both compounds at ambient pressure is continuous (a second order transition).

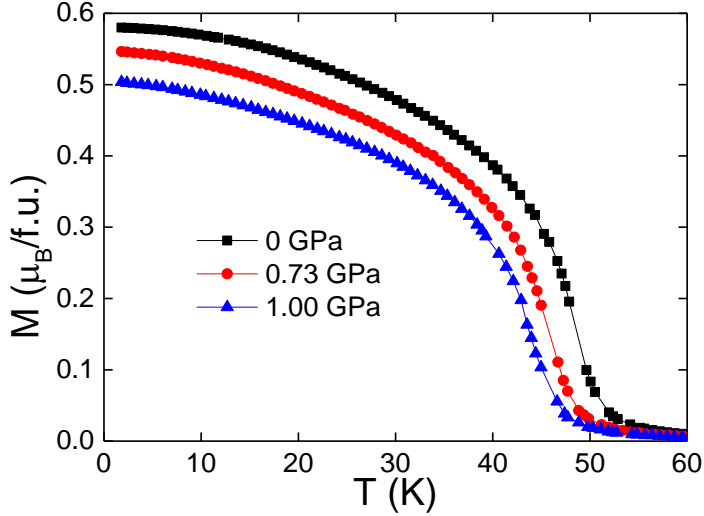

*Fig. 1:* Temperature dependence of the magnetization of UCoGa at different pressures, in an applied field of 0.1 T.

The magnetization isotherms of UCoGa and URhGa were measured at 1.8 K, at the same pressures as the corresponding temperature dependencies of thermomagnetic curves (Fig. 3).

The values of spontaneous magnetization ($M_s$) were obtained by extrapolating the parts (above 1 T in order to avoid effects related to domains and superconducting phase of Pb) of magnetization curves to zero magnetic field. The $M_s$ values obtained for UCoGa and URhGa at different pressures are listed in Table 1. In the case of UCoGa $M_s$ is clearly decreasing with increasing pressure, whereas the $M_s$ of URhGa decreases only slightly between 0 and 0.6 GPa, and remains unchanged with higher pressures up to 1 GPa.

The Arrot plots of UCoGa and URhGa magnetization data, for all pressure points, are shown in Fig. 4. The slopes of the plots determined by linear regression and the values of $M_s$ and $T_C$ were used to determine values of the TSFT parameters $F_1$, $T_A$, $T_0$ and $T_C/T_0$ at different pressure points, and are shown in Table 1 and Fig. 5.

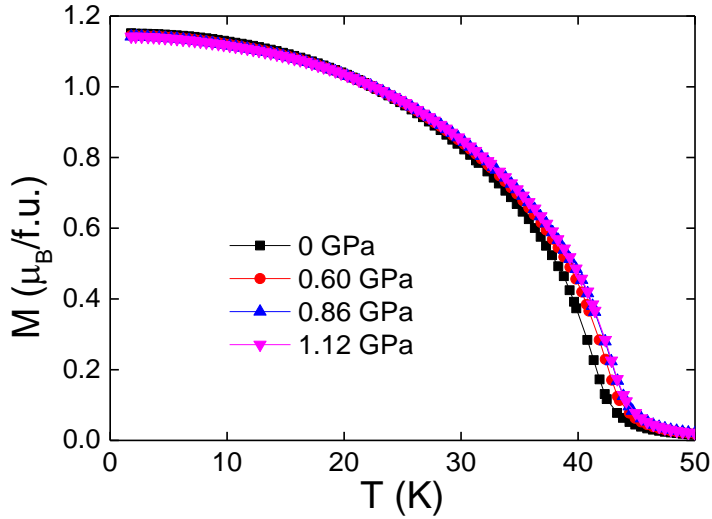

*Fig. 2:* Temperature dependence of the magnetization of URhGa at different pressures, in an applied field of 0.1 T.

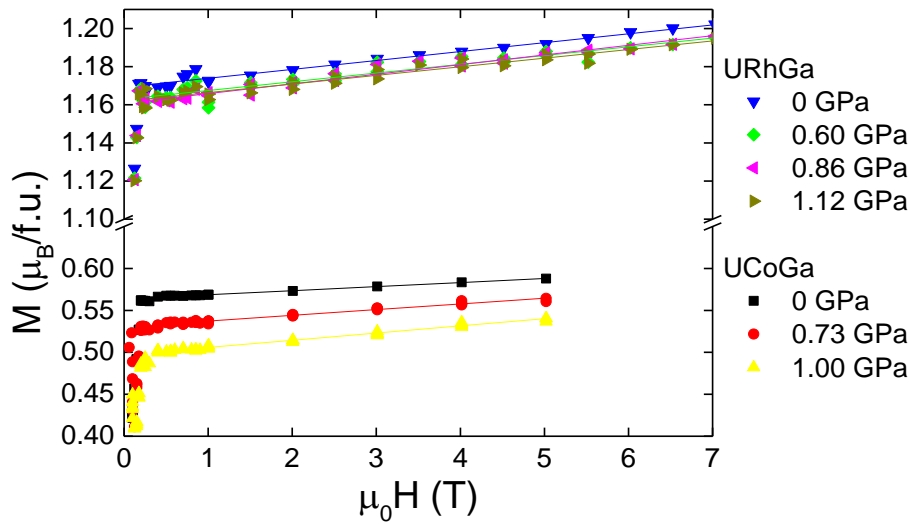

*Fig. 3:* Magnetization curves of UCoGa and URhGa measured at 1.8 K in magnetic fields applied along the *c*-axis at various pressures with contribution from pressure cell already subtracted. The contribution from the pressure cell was determined by comparing data in 0 GPa and ambient pressure. Lines are used as guide to eye.

To facilitate the discussion of these parameters, and their development with pressure, the complexity of the role of U 5*f* electrons on the electronic structure and magnetism should be taken into account. The variable dual character of the 5*f* electrons of U ions (partially localized, partially itinerant) [13–15] found in various crystallographic and chemical environments in U compounds, is reflected in the wide range of their observed magnetic behaviors. The 5*f* wave functions are widely extended in space and allow strong 5*f*-ligand hybridization in the compounds, which destroys the original atomic character of the 5*f* wave functions and the related magnetic moments. In the strong hybridization limit, the 5*f* magnetic moments vanish and magnetic order is lost. The 5*f*-ligand hybridization, however, may also enforce the magnetic ordering because it mediates the indirect exchange interaction of pairs of 5*f*-electron magnetic moments born at U ions via the involved ligand [1]. In materials in which the 5*f*-ligand hybridization is not too strong, the 5*f*-moments remain stable and increasing hybridization enhances the exchange interaction and causes an increase in $T_C$. The 5*f*-ligand hybridization originates from overlaps of the U 5*f* –wave functions, with the wave functions of ligand valence-electrons and depends critically on the distances between involved ions. Thanks to compressibility, interatomic distances, and consequently the 5*f*-ligand hybridizations, can be controlled by external pressure. Isostructural families of materials, such as U*TX* compounds with a hexagonal ZrNiAl structure, provides a useful playground for testing the effect of ligand species and U-ligand interatomic distances on the degree of 5*f*-electron delocalization and their implications for critical parameters of magnetic ordering while maintaining constant crystallographic symmetry. UCoGa and URhGa were selected for our study because they represent two groups of U*TX* compounds characterized by different signs of pressure effect on $T_C$, quantitatively expressed by $d\ln T_C/dP$ (see Table 1). The two compounds have similar values of Curie temperature ($T_C$ = 48.8 and 41.1 K, respectively) but the low-temperature spontaneous magnetization $M_s$ = 1.16 $\mu_B$/f.u. of URhGa is more than double the $M_s$ = 0.56 $\mu_B$/f.u. of UCoGa. The strongly reduced U moment in UCoGa is a clear indication of a much stronger delocalization of 5*f*-electrons, compared to URhGa. Such a situation can be intuitively expected when we take into account the much smaller lattice parameters, in particular the *a*, in UCoGa with respect to URhGa. Consequently, the corresponding U-U and U-*T* interatomic distances within the basal plane imply a much larger 5*f*-5*f* and 5*f*-3*d* wave-function overlaps, with stronger 5*f*-ligand hybridization leading to much more delocalized 5*f*-electrons in UCoGa.

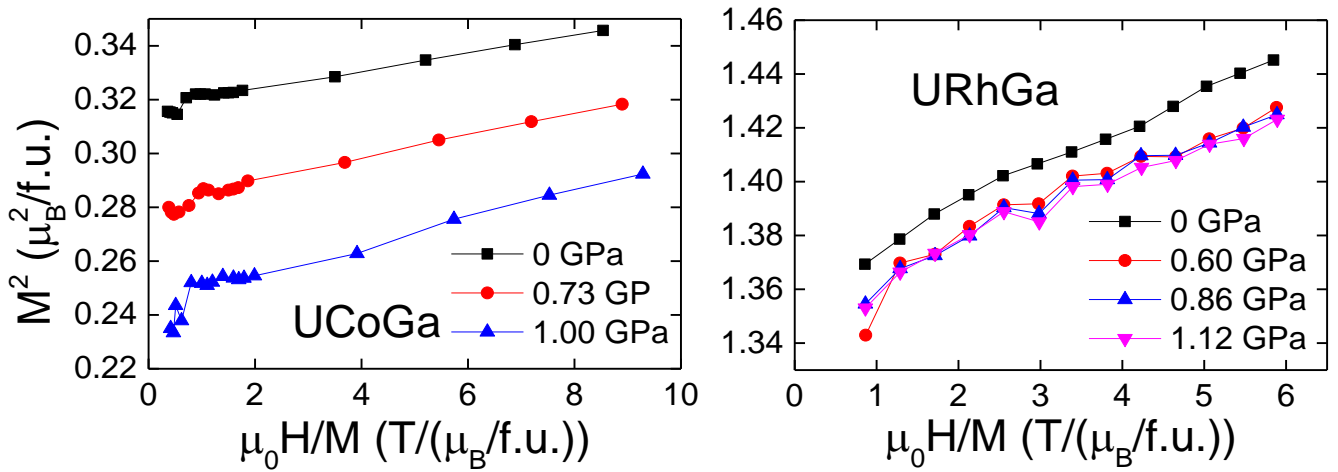

*Fig. 4*: The Arrot plots of magnetization isotherm data UCoGa and URhGa measured at different pressures.

From the point of view of TSFT, this situation is clearly reflected in the significantly higher magnitude of the $T_C/T_0$ ratio (= 0.495) obtained for URhGa compared to the $T_C/T_0$ = 0.182 obtained for UCoGa. A higher $T_C/T_0$ ratio represents more localized magnetic electrons. Consistently with better localization of the 5*f* electrons, the SF spectra for URhGa are much narrower both in energy and wave-vector space (smaller $T_0$ and $T_A$) than for UCoGa.

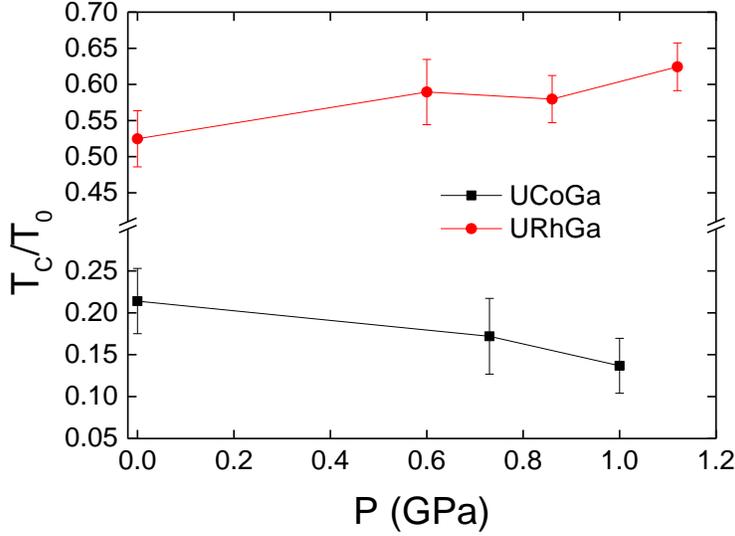

*Fig. 5:* Pressure dependence of degree of localization, $T_C/T_0$, for UCoGa and URhGa

In Table 1 we can also see that the application of hydrostatic pressure on UCoGa and URhGa has a different effect on the corresponding values of $M_s$, $T_C$ and the TSFT parameters. With increasing applied pressure on UCoGa, $M_s$, $T_C$ and $T_C/T_0$ decrease rapidly. $T_0$ increases whereas $T_A$ almost doesn't change. On the other hand, $T_C$, $T_C/T_0$ and $T_A$ of URhGa slightly increase, whilst $M_s$ and $T_0$ remain invariant.

These results fit well with the general scenario of relations between electronic structure and magnetism in U compounds, discussed above. URhGa appears in the conditions of moderate hybridization when a pressure-induced increase of hybridization increases the exchange of an interaction leading to an increase of $T_C$ without visible effect on $M_s$. On the contrary, UCoGa is in a strong hybridization mode, where a significant suppression of the U magnetic moments due to the increasing pressure prevails over the increased exchange integral, so that the $T_C$ and $T_C/T_0$ decrease rapidly with increasing pressure. From the point of view of TSFT, a decrease of $T_C/T_0$ means an increasing itinerancy of 5*f* electrons, which leads to an extension of the SF spectrum indicated by an increasing $T_0$ value.

*Table 1:* Experimental values of $T_C$ and $M_s$ with calculated values of $F_1$, $T_0$, $T_A$ and $T_C/T_0$ from TSFT for UCoGa and URhGa at different pressures.

|  | $P$ (GPa) | $T_C$ (K) | $M_s$ ($\mu_B$/f.u.) | $F_1$ (K) | $T_A$ (K) | $T_0$ (K) | $T_C/T_0$ |
|---|---|---|---|---|---|---|---|
| UCoGa | 0 | 48.8 | 0.56 | 3060 | 1750 | 267 | 0.182 |
|  | 0.73 | 46.0 | 0.53 | 2320 | 1700 | 330 | 0.139 |
|  | 1 | 44.5 | 0.50 | 1880 | 1720 | 424 | 0.105 |
| URhGa | 0 | 41.1 | 1.17 | 741 | 480 | 83 | 0.495 |
|  | 0.6 | 42.2 | 1.16 | 776 | 498 | 85 | 0.496 |
|  | 0.86 | 42.6 | 1.16 | 824 | 508 | 83 | 0.512 |
|  | 1.12 | 42.8 | 1.16 | 853 | 513 | 82 | 0.521 |

*Table 2:* Evolution of room-temperature lattice parameters, ambient-pressure experimental values of $T_C$ and $M_s$, calculated values of $F_1$, $T_0$, $T_A$ and $T_C/T_0$ from TSFT and initial pressure coefficients $\partial \ln T_C/\partial P$ values for $P \to 0$ within a group of isostructural U$TX$ ferromagnets crystallizing in the hexagonal ZrNiAl-type structure.

| $a$ (pm) | $c$ (pm) | Compound | $T_C$ (K) | $M_s$ ($\mu_B$/f.u.) | $F_1$ (K) | $T_A$ (K) | $T_0$ (K) | $T_C/T_0$ | $\partial \ln T_C/\partial P$ GPa$^{-1}$ |
|---|---|---|---|---|---|---|---|---|---|
| 669.1 | 396.6 | UCo$_{0.98}$Ru$_{0.02}$Al | 22.7 | 0.36 | 2311 | 1540 | 274 | 0.083 | - 1.074 |
| 669.3 | 393.3 | UCoGa [16] | 48.8 | 0.56 | 3060 | 1750 | 267 | 0.182 | - 0.090 |
| 695.8 | 401.4 | UIrAl [17,18] | 62 | 0.96 | 820 | 861 | 241 | 0.257 | - 0.005 |
| 697.0 | 402.1 | URhAl [16,18–20] | 26.2 | 1.05 | 428 | 340 | 64.5 | 0.365 | - 0.003 |
| 700.6 | 394.5 | URhGa [16] | 41.1 | 1.17 | 741 | 480 | 83 | 0.495 | 0.037 |
| 701.7 | 412.4 | UPtAl [16,18,21-23] | 43.5 | 1.38 | 615 | 395 | 67.8 | 0.642 | 0.059 |

These are the signatures that URhGa gradually moves towards the strong 5$f$-ligand hybridization regime where the washout of U magnetic moments dominates due to increasing itinerancy of 5$f$-electrons, so that $T_C$ and also $T_C/T_0$ will decrease in higher pressures. In Table 2, UCoGa and URhGa are compared with several other isostructural U$TX$ ferromagnets for which the pressure effects are known. To facilitate further discussion also the lattice parameters are displayed.

When inspecting Table 2 closely, one can observe a clear relation between the evolution of the lattice parameter $a$, the spontaneous magnetization $M_s$, the $T_C/T_0$ ratio, representing in TSFT the degree of 5$f$-electron localization and the d$T_C$/d$P$ values. These parameters increase monotonously throughout Table 2. The increasing $a$ is directly associated with the proportionally increasing U-U and U-$T$ interatomic distances within the basal plane around which most of the 5$f$-5$f$ overlaps and 5$f$-$d$ hybridizations occur. Increasing localization of 5$f$ electrons can be expected when $a$ increases, which correlates with the increasing $T_C/T_0$ ratio.

As to the sign of the initial pressure response of $T_C$, Table 2 has two parts: a) URhAl and compounds above it with $\partial \ln T_C/\partial P < 0$ and b) compounds with $\partial \ln T_C/\partial P > 0$ (URhGa, UPtAl). The limit $T_C/T_0$ value separating materials with positive to negative pressure response of $T_C$ respectively, can be roughly estimated as 0.43.

The crossover of the change of $\partial \ln T_C/\partial P$ sign is located between URhAl and URhGa. The opposite signs of d$T_C$/d$P$ values observed for URhGa (positive) and URhAl (negative) demonstrate that the 5$f$-3$p$ (URhAl) hybridization causes a stronger delocalization of the 5$f$-electrons than the 5$f$-4$p$ (URhGa) hybridization.

URhAl shows an initial slight negative d$T_C$/d$P$. The rate of $T_C$ decrease accelerates with increasing pressure towards the loss of ferromagnetism observed at a tricritical point at ~ 5.2 GPa [4] (similar to UCoGa [5]).

Investigation of URhGa to higher pressures [7] revealed $T_C$ in URhGa increases linearly (d$T_C$/d$P$ ~ 1.1 K/GPa) whereas d$M_s$/d$P$ decreases slightly (d$M_s$/d$P$ ~ - 0.02 $\mu_B$/f.u./GPa) with increasing pressure up to 4 GPa. In higher pressures, d$T_C$/d$P$ gradually decreases and the pressure, where $T_C$ reaches the maximum value can be expected somewhere between 6 to 9 GPa. $M_s$ decreases much faster with increasing pressure above 4 GPa (d$M_s$/d$P$ ~ -0.08 $\mu_B$/f.u./GPa).

UPtAl also exhibits an increasing $T_C$ with increasing pressure [6,22,23]. $T_C$ reaches the maximum value at 6 GPa and then decreases with further increasing pressure up to ~ 17 GPa, where the ferromagnetism is suppressed.

UCoAl, which belongs to the discussed family of isostructural compounds, deserves to be mentioned here, too, although it is not ferromagnetic at ambient conditions. It is a paramagnet undergoing a metamagnetic transition in a field applied along the *c*-axis [3,24]. A slight negative chemical pressure accomplished by substituting, e.g. 4% of Lu for U, induces a ferromagnetic ground state [25] whereas application of hydrostatic pressure pushes the metamagnetic transition to higher fields. The study of UCo(Al,Ga) solid solutions [26] revealed the gradual transformation from a paramagnetic ground state of UCoAl to ferromagnetism in UCoGa, with the onset of ferromagnetism around 20% Ga. Detailed investigation on single crystals of selected $UCoAl_{1-x}Ga_x$ compositions is desirable to test the TSFT applied to the evolution of itinerancy of 5*f*-electron states in the vicinity of the critical Ga concentration for the onset of ferromagnetism.

So far the discussion was considering experiments involving effects induced by applying hydrostatic pressure. When increasing uniaxial stress is applied on the UCoAl single crystal along the hexagonal *c*-axis a decrease of the critical field of metamagnetic transition ($H_c$) is observed until it vanishes and a ferromagnetic ordering is established [27,28]. On the other hand, the UCoAl crystal is compressed along *a*, an increase in $H_c$ is observed [29,30].

These results can be understood when we realize that the material exerted to hydrostatic pressure is compressed and interatomic distances in different directions reduced according to the respective linear compressibilities. Uniaxial stress is pushing together atoms along the stress direction whereas the atoms in the perpendicular plane move apart. In the hexagonal U*TX* compounds adopting the ZrNiAl-type structure the most of the 5*f*-5*f* overlaps and 5*f*-*d* hybridizations occur around the basal plane which is perpendicular to the *c*-axis. The *c*-axis uniaxial pressure then leads to better localization of 5*f*-electrons and hitherto the stabilization of U magnetic moments. As a result, $H_c$ decreases with increasing uniaxial pressure and finally ferromagnetism emerges. This process is boosted by the fact that the compressibility along the *a*-axis is, in this family of compounds, usually several times (~ 3 times in UCoAl) larger than that along the *c*-axis [22,31].

It is should be noted that the above mentioned *a*-axis stress experiment [29,30] in the hexagonal UCoAl crystal cannot bring definitive results with respect to this topic. A basal-plane stress experiment, in which the basal plane of the UCoAl crystal is uniformly stressed in all directions whereas the crystal is free to dilate within the *c*-axis direction, is desirable for affecting the 5*f*-5*f* overlaps and 5*f*-*d* hybridizations around the plane.

Some of U*TX* ferromagnets, namely UCoGe, URhGe and URhSi crystallize in the orthorhombic structure of the TiNiSi-type. The first two of them are the well-known ambient-pressure ferromagnetic superconductors [32,33]. This group of compounds allows us to investigate the extent to which the $T_C/T_0$ - $\partial \ln T_C/\partial P$ correlation observed in ZrNiAl-type hexagonal family compounds is valid in isoelectronic U*TX* compounds crystallizing in a lower symmetric structure. In Table 3, they are sorted according to increasing values of $\partial \ln T_C/\partial P$.

UCoGe, one of the weakest itinerant 5*f*-electron ferromagnets UCoGe shows a very small $T_C/T_0$ value, which is by more than one order of magnitude lower than the value for $UCo_{0.98}Ru_{0.02}Al$. The $T_C/T_0$ value for URhSi is comparable to the $T_C/T_0$ value for UCoGa. Similar to $UCo_{0.98}Ru_{0.02}Al$ and UCoGe the Curie temperatures of UCoGe and URhSi decrease with applying hydrostatic pressure. The rate of decrease is four times faster in the case of UCoGe that in URhSi. For UCoGe and URhSi the $T_C/T_0$ - $\partial \ln T_C/\partial P$ correlation seems to be valid.

URhGe, however, is an obvious exception. Although it shows lower $M_s$, $T_C$ and also $T_C/T_0$ values than URhSi, the response of $T_C$ of URhGe to hydrostatic pressure is positive.

At this point, it is interesting to note that when uniaxial stress is applied on URhGe along the *b*-axis, which is perpendicular to the easy-magnetization direction (*c*-axis), the Curie temperature decreases at a rate $\partial n T_C/\partial P \sim -0.17$ GPa$^{-1}$ and simultaneously the critical temperature of superconductivity increases [34].

*Table 3: Evolution of lattice parameters, ambient-pressure experimental values of $T_C$ and $M_s$, calculated values of $F_1$, $T_0$, $T_A$ and $T_C/T_0$ from TSFT and initial pressure coefficients $\partial n T_C/\partial P$ values for $P \to 0$ within a group of isostructural UTX ferromagnets crystallizing in the orthorhombic TiNiSi-type-type structure.*

| a (pm) | b (pm) | c (pm) | Compound | $T_C$ (K) | $M_s$ ($\mu_B$/f.u.) | $F_1$ (K) | $T_0$ (K) | $T_A$ (K) | $T_C/T_0$ | $\partial \ln T_C/\partial P$ GPa$^{-1}$ |
|---|---|---|---|---|---|---|---|---|---|---|
| 684.66 | 420.65 | 722.74 | UCoGe [18,35,36] | 2.4 | 0.039 | 28700 | 362 | 5920 | 0.007 | -0.58 |
| 702.3 | 412.1 | 745.8 | URhSi [18,37,38] | 10.5 | 0.571 | 520 | 64.5 | 354 | 0.163 | ~ -0.2 |
| 689.79 | 434.03 | 753.58 | URhGe [18,35,39] | 9.5 | 0.41 | 1100 | 78.4 | 568 | 0.121 | +0.13 |

In Table 4 relevant data are shown for the other two U intermetallic compounds (UGe$_2$, UGa$_2$) for which sufficient information for the discussion in this paper is available. The first reported uranium ferromagnetic superconductor, orthorhombic UGe$_2$ [40,41] represents another striking example for which the above discussed $T_C/T_0$ - $\partial n T_C/\partial P$ correlation is not observed. It exhibits one of the highest values of spontaneous magnetization among U intermetallics and quite high $T_C/T_0$, both much higher than found for URhGe, but contrary to URhGe it shows a fast decrease of $T_C$ when exerted to hydrostatic pressure.

On the other hand the hexagonal compound UGa$_2$, [42,43] is probably an example of a uranium intermetallic compound with the most localized 5f-electrons. It has a spontaneous magnetic moment of almost 3 $\mu_B$/U and $T_C/T_0$ larger than one. Consistent with expectations, the $T_C$ of UGa$_2$ increases when the compound is exposed to hydrostatic pressure.

*Table 4: Ambient-pressure experimental values of $T_C$ and $M_s$, calculated values of $F_1$, $T_0$, $T_A$ and $T_C/T_0$ from TSFT and initial pressure coefficients $\partial n T_C/\partial P$ values for $P \to 0$ in UGe$_2$ and UGa$_2$.*

| Compound | $T_C$ (K) | $M_s$ ($\mu_B$/f.u.) | $F_1$ (K) | $T_0$ (K) | $T_A$ (K) | $T_C/T_0$ | $\partial \ln T_C/\partial P$ GPa$^{-1}$ |
|---|---|---|---|---|---|---|---|
| UGe$_2$ [18,40] | 52.6 | 1.41 | 554 | 92.2 | 442 | 0.571 | -0.13 |
| UGa$_2$ [18,42,43] | 123 | 2.94 | 273 | 94.8 | 311 | 1.12 | 0.03 |

The opposite pressure response of URhGe and UGe$_2$ Curie temperatures with respect to expectations from the $T_C/T_0$ - $\partial n T_C/\partial P$ correlation observed for the uniaxial U*TX* ferromagnets adopting the hexagonal ZrNiAl-type structure indicates that the correlation is not a universal feature of U ferromagnets independent from the underlying crystal structure. The reason may be connected with a more complex anisotropy of spin fluctuations in compounds with lower symmetries.

The compression of the 5f-electron density towards the basal plane, together with strong spin-orbit coupling in the hexagonal U*TX* compounds with the ZrNiAl-type structure are the sources of huge uniaxial magnetocrystalline anisotropy with the easy-magnetization axis along the hexagonal c-axis. This makes the magnetism in these materials almost two dimensional. The negligible magnetization observed in the basal plane (perpendicular to the c-axis) corroborates the idea that the transversal spin fluctuations are negligible [44] and the spin-fluctuation behavior is simplified. The complex magnetocrystalline anisotropy in the orthorhombic U$TX$ compounds and UGe$_2$ with different components along the main crystallographic directions together with the anisotropy of compressibility [45] and thermal expansion [46–48] probably does not allow us to apply the model applicable to the hexagonal U$TX$ compounds on these materials.

Further investigation of magnetoelastic behavior of a wider range of U intermetallics is desirable for collecting sufficient amount of information needed for testing possible theoretical approaches.

## Conclusions

The magnetization of UCoGa and URhGa has been measured as a function of temperature, magnetic field and hydrostatic pressure. The results were analyzed in the framework of Takahashi's spin-fluctuation theory. The TSFT parameters $T_0$ and $T_A$ characterizing the distribution widths of the SF spectrum in the energy and wave vector space, respectively, and $T_C/T_0$ characterizing the degree of the 5$f$-electron localization, have been determined for hydrostatic pressures up to 1 GPa. Examination of available experimental data for U$TX$ ferromagnets adopting the hexagonal ZrNiAl-type structure, revealed correlations between the degree of 5$f$-electron localization, represented by $T_C/T_0$ parameters, and the response of the Curie-temperature on the applied pressure. The $T_C/T_0$ value of 0.43 has been estimated as the limit separating compounds with lower (higher) $T_C/T_0$ values and $\partial \ln T_C/\partial P < 0$ ($\partial \ln T_C/\partial P > 0$). Comparison with corresponding data known for the orthorhombic U$TX$ ferromagnets and UGe$_2$ indicates that the above mentioned correlation does not have universal validity for U ferromagnets outside the ZrNiAl-structure U$TX$ family.

## Acknowledgements

Experiments were performed in MGML (www.mgml.eu), which is supported within the program of Czech Research Infrastructures (project no. LM2018096). We would also like to thank to Dr. Ross Colman for proofreading of the text, and language corrections and Dr. Martin Míšek for measuring the pressure coefficient of UCo$_{0.98}$Ru$_{0.02}$Al.